\documentclass [prl,aps,letterpaper,twocolumn,superscriptaddress,amsmath,amssymb,floatfix] {revtex4}

\usepackage{graphicx}
\usepackage{textcomp}
\usepackage{epstopdf}

\newcommand{\bra}[1]{\ensuremath{\left\langle#1\right|}}
\newcommand{\ket}[1]{\ensuremath{\left|#1\right\rangle}}

\begin{document}

\title{Tracking Photon Jumps with Repeated Quantum Non-Demolition Parity Measurements}
\author{L.~Sun\footnote{current address: Center for Quantum Information, Institute for Interdisciplinary Information Sciences, Tsinghua University, Beijing, P. R. China}}
\affiliation{Departments of Applied Physics and Physics, Yale University, New Haven, CT 06511, USA}
\author{A.~Petrenko}
\affiliation{Departments of Applied Physics and Physics, Yale University, New Haven, CT 06511, USA}
\author{Z.~Leghtas}
\affiliation{Departments of Applied Physics and Physics, Yale University, New Haven, CT 06511, USA}
\author{B.~Vlastakis}
\affiliation{Departments of Applied Physics and Physics, Yale University, New Haven, CT 06511, USA}
\author{G.~Kirchmair\footnote{current address: Institut f\"{u}r Experimentalphysik, Universit\"{a}t Innsbruck, Technikerstra{\ss}e 25, A-6020 Innsbruck, Austria;
Institut f\"{u}r Quantenoptik und Quanteninformation, \"{O}sterreichische Akademie der Wissenschaften, Otto-Hittmair-Platz 1, A-6020 Innsbruck, Austria}}
\affiliation{Departments of Applied Physics and Physics, Yale University, New Haven, CT 06511, USA}
\author{K.~M.~Sliwa}
\affiliation{Departments of Applied Physics and Physics, Yale University, New Haven, CT 06511, USA}
\author{A.~Narla}
\affiliation{Departments of Applied Physics and Physics, Yale University, New Haven, CT 06511, USA}
\author{M.~Hatridge}
\affiliation{Departments of Applied Physics and Physics, Yale University, New Haven, CT 06511, USA}
\author{S.~Shankar}
\affiliation{Departments of Applied Physics and Physics, Yale University, New Haven, CT 06511, USA}
\author{J.~Blumoff}
\affiliation{Departments of Applied Physics and Physics, Yale University, New Haven, CT 06511, USA}
\author{L.~Frunzio}
\affiliation{Departments of Applied Physics and Physics, Yale University, New Haven, CT 06511, USA}
\author{M.~Mirrahimi}
\affiliation{Departments of Applied Physics and Physics, Yale University, New Haven, CT 06511, USA}
\affiliation{INRIA Paris-Rocquencourt, Domaine de Voluceau, B.P.~105, 78153 Le Chesnay Cedex, France}
\author{M.~H.~Devoret}
\affiliation{Departments of Applied Physics and Physics, Yale University, New Haven, CT 06511, USA}
\author{R.~J.~Schoelkopf}
\affiliation{Departments of Applied Physics and Physics, Yale University, New Haven, CT 06511, USA}
 \maketitle

\textbf{Quantum error correction (QEC) is required for a practical quantum computer because of the fragile nature of quantum information~\cite{Nielsen}. In QEC, information is redundantly stored in a large Hilbert space and one or more observables must be monitored to reveal the occurrence of an error, without disturbing the information encoded in an unknown quantum state. Such observables, typically multi-qubit parities such as  $\left<\sigma_{1}^{x}\sigma_{2}^{x}\sigma_{3}^{x}\sigma_{4}^{x}\right>$, must correspond to a special symmetry property inherent to the encoding scheme. Measurements of these observables, or error syndromes, must also be performed in a quantum non-demolition (QND) way and faster than the rate at which errors occur. Previously, QND measurements of quantum jumps between energy eigenstates have been performed in systems such as trapped ions~\cite{Bergquist, Sauter, Nagourney}, electrons~\cite{Peil}, cavity quantum electrodynamics (QED)~\cite{Gleyzes, Guerlin}, nitrogen-vacancy (NV) centers~\cite{Jelezko, Robledo}, and superconducting qubits~\cite{Vijay, HatridgeShankar}. So far, however, no fast and repeated monitoring of an error syndrome has been realized. Here, we track the quantum jumps of a possible error syndrome, the photon number parity of a microwave cavity, by mapping this property onto an ancilla qubit. This quantity is just the error syndrome required in a recently proposed scheme for a hardware-efficient protected quantum memory using Schr\"{o}dinger cat states in a harmonic oscillator~\cite{Leghtas2}. We demonstrate the projective nature of this measurement onto a parity eigenspace by observing the collapse of a coherent state onto even or odd cat states. The  measurement is fast compared to the cavity lifetime, has a high single-shot fidelity, and has a 99.8\% probability per single measurement of leaving the parity unchanged. In combination with the deterministic encoding of quantum information in cat states realized earlier~\cite{Leghtas1,Vlastakis}, our demonstrated QND parity tracking represents a significant step towards implementing an active system that extends the lifetime of a quantum bit.}

\begin{figure}[b]
\includegraphics{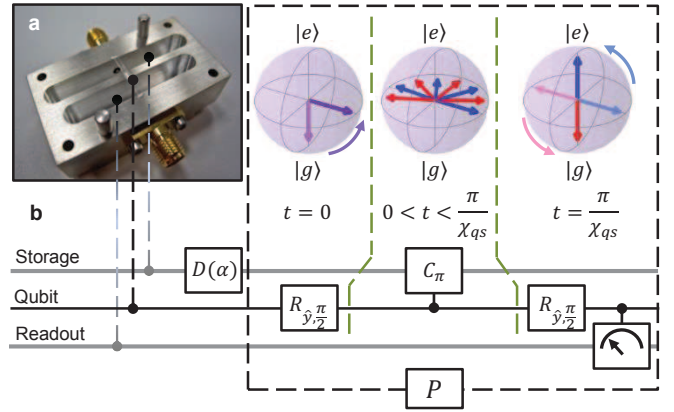}
\caption {Experimental device and parity measurement protocol (P) of a photon state. (a) Bottom half of the device containing a transmon qubit located in a trench and coupled to two waveguide cavities. The low frequency cavity, with $\omega_s/2\pi=7.216$ GHz and a lifetime of $\tau_0=55~\mu$s, is used to store and manipulate quantum states. The high frequency cavity, with $\omega_m/2\pi=8.174$ GHz and a lifetime of 30~ns, allows for a fast readout of the qubit. (b) The protocol for measuring the parity of the storage cavity field. After an initial coherent displacement of $\alpha$, a Ramsey-type measurement is performed. It consists of two $\pi/2$ pulses separated by $t=\pi/\chi_{qs}$, followed by a projective measurement of the qubit, where $\chi_{qs}$ is the dispersive interaction between the qubit and the storage cavity. In this schematic, with the qubit initially in the ground state $\ket{g}$, the Ramsey-type measurement maps the even (odd) photon state onto the $\ket{e}$ ($\ket{g}$) state of the qubit. A subsequent projective measurement indicates the cavity state parity. The second $\pi/2$ pulse can be either $R_{\hat{y},-\frac{\pi}{2}}$ or $R_{\hat{y},\frac{\pi}{2}}$, simply switching the interpretation of the result of the qubit measurement.}
\end{figure}

Besides their necessity in quantum error correction and quantum information, QND measurements play a central role in quantum mechanics. The application of an ideal projective QND measurement yields a result corresponding to an eigenvalue of the measured operator, and projects the system onto the eigenstate associated with that eigenvalue. Moreover, the measurement must leave the system in that state, so that subsequent measurements always return the same result. The hallmark of a continuously repeated high fidelity QND measurement is that it demonstrates a canonical \textit{Gedankenexperiment}: individual quantum jumps between eigenstates are resolved in time on a single quantum system. This ideal measurement capability has only been experimentally realized in the last few decades. The jumps of a two-level system (qubit) between its energy eigenstates were first observed for single trapped ions~\cite{Bergquist, Sauter, Nagourney}, and later in single NV centers in diamond~\cite{Jelezko, Robledo}. The jumps of an oscillator between eigenstates with different numbers of excitations (Fock states), were first observed for the motion of an electron in a Penning trap~\cite{Peil}. More recently, the observation of quantum jumps of light in cavity QED~\cite{Gleyzes, Guerlin}, where the number of microwave photons in a cavity is probed with Rydberg atoms, has enabled a range of new experiments in quantum feedback and control~\cite{Deleglise, Sayrin}. 

\begin{figure}
\includegraphics{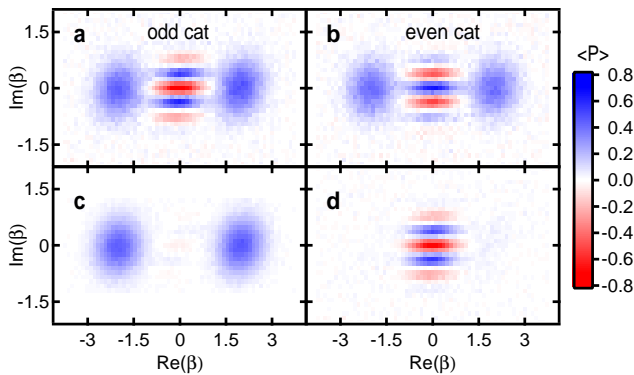}
\caption {Ensemble averaged Wigner functions of cat states in the cavity created by single-shot parity measurements of an initial coherent state in the cavity. The Wigner functions are mapped out with varying displacements $\beta$~\cite{Haroche}. Here we follow the protocol depicted in Fig.~1b, using a $R_{\hat{y},\frac{\pi}{2}}$ as the second pulse. The qubit is always initialized to the $\ket{g}$ state through post selection on an initial measurement. (a) Odd cat by post selection on the $\ket{g}$ states. (b) Even cat by post selection on the $\ket{e}$ states. (c) No post selection of the parity measurement, thus tracing over the qubit state. Fringes disappear, indicating a mixed state of two coherent states. (d) The normalized difference $(\frac{\rm{a}-\rm{b}}{2})$, or the expectation of the parity weighted by $\left<\sigma_{z}\right>$ of the ancilla, emphasizing the interference fringes.}
\end{figure}   

An analogous system to cavity QED is the combination of microwave photons in a superconducting resonator with superconducting qubits, known as circuit QED~\cite{DevoretSchoelkopf}. The strong dispersive interaction of qubit and photon, as in Rydberg atom cavity QED, allows either the qubit or cavity to act as a QND probe of the other component. With the advent of quantum-limited parametric amplifiers~\cite{Castellanos,Bergeal,Vijay}, measurement techniques for superconducting devices have rapidly advanced. For instance, the frequency shift of a cavity has been recently used  to observe the quantum jumps of a qubit between energy eigenstates~\cite{Vijay, HatridgeShankar}. So far, however, there have been no observations of jumps for the cavity field in circuit QED. 

In this work, we use the dispersive qubit-cavity interaction of circuit QED to observe the jumps of photon number parity. Importantly, these jumps reveal the loss of individual photons without  projecting the system onto a state of definite number or energy, but rather into an eigenspace of even or odd photon number. This characteristic is a crucial requirement for future applications in quantum information, where the parity measurement serves as the error syndrome for correcting a quantum memory. Even in the presence of rapidly repeated measurements, the smooth decay of the ensemble averaged parity is largely unperturbed. However, when individual time records of the measurement are examined, the parity is observed to take on only the extremal values, $\pm 1$, indicating the projective nature of each individual measurement. On examining the statistics of the jumps recorded over many trajectories, we find excellent agreement with a numerical simulation, suggesting 85\% of the jumps for states with an average photon number $\bar{n}=4$ are faithfully detected (see Supplementary Material C). When selecting on the outcome of a single parity measurement, we observe via Wigner tomography~\cite{Haroche} the creation of cat states with $\bar{n}$ up to 4.

\begin{figure*}
\centering
\includegraphics{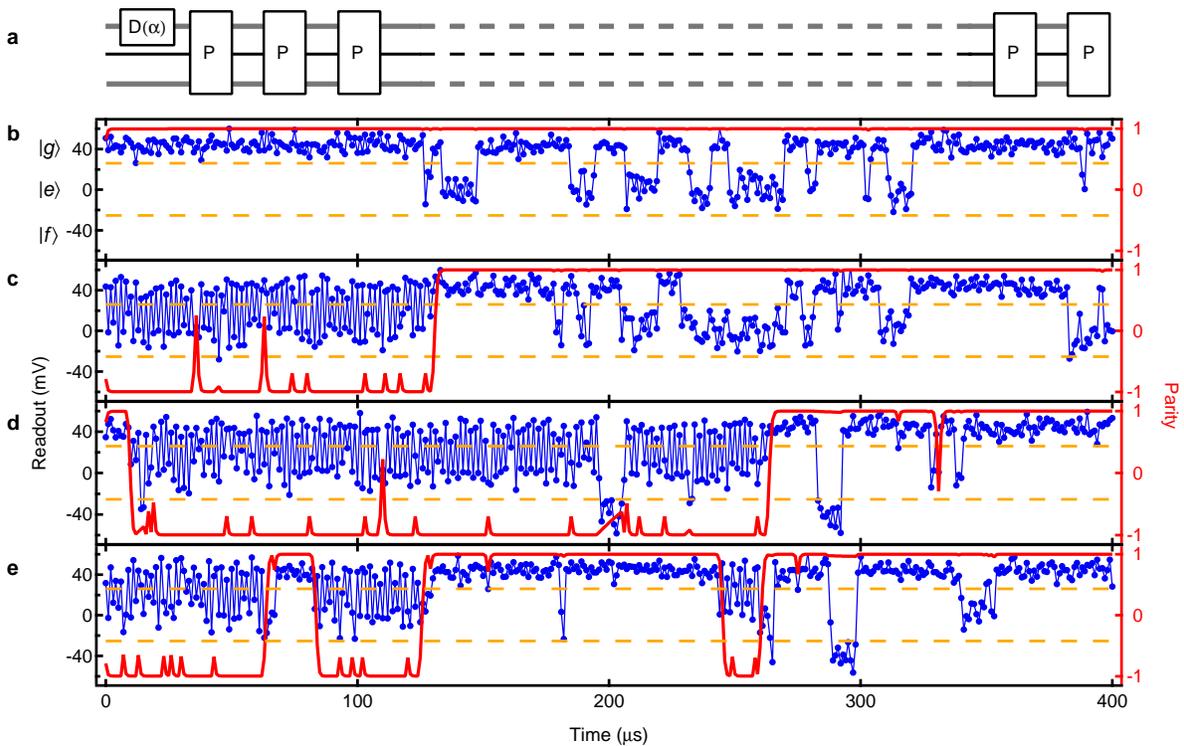}
\caption {Typical repeated single-shot parity measurement traces revealing photon jumps in real time. Horizontal dashed orange lines delineate the thresholds to distinguish $\ket{g}, \ket{e}$, and higher excited states of the qubit, denoted as $\ket{f}$. The red traces show the quantum filter that best estimates the parity at every point. The filter has a finite response time and thus does not trust that a brief change in the measurement pattern corresponds to an actual parity jump. (a) In this protocol we switch the sign of the second pulse, using a $R_{\hat{y},-\frac{\pi}{2}}$ instead of a $R_{\hat{y},\frac{\pi}{2}}$. The repetition time of the parity measurement is 1~$\mu$s, and traces b-e all have an initial displacement of $|\alpha|=1$. (b) For the most part the correlation between neighboring measurements is positive, indicating an even parity state for the whole 400 $\mu$s. The changes in the qubit state between 120~$\mu$s and 320~$\mu$s are likely due to qubit decoherence during the parity measurement. (c) One parity jump is observed by the change in the measurement pattern (oscillating vs. constant) at about 130~$\mu$s. (d) Two parity jumps are recorded at about 10~$\mu$s and then again at 260~$\mu$s. The change of pattern at about 200 $\mu$s is a result of the qubit leaving the computational space for higher excited states, a feature that disables the parity measurement until the qubit returns to either $\ket{g}$ or $\ket{e}$. (e) A trace with all features described above included.  In this particular trajectory, the filter can clearly resolve five photon jump events.}
\end{figure*} 

In our experiment, we employ a three-dimensional circuit QED architecture~\cite{Paik} with a single superconducting transmon qubit coupled to two waveguide cavities~\cite{Kirchmair,Vlastakis}, as shown in Fig.~1a.  Our qubit has a transition frequency of $\omega_q/2\pi=5.938$ GHz, an energy relaxation time $T_1=8~\mu$s, and a Ramsey time $T_2^*=5~\mu$s. The high frequency cavity, with $\omega_m/2\pi=8.174$ GHz and a lifetime of 30~ns, serves only as a fast readout of the qubit state. In order to perform a high-fidelity single-shot dispersive readout of the qubit, we use a Josephson bifurcation amplifier (JBA) operating in a double-pumped mode~\cite{Kamal,Murch,Sliwa} as the first stage of amplification. The low frequency cavity, with $\omega_s/2\pi=7.216$ GHz and a lifetime of $\tau_0=55~\mu$s, stores the photon states which are measured and manipulated. Exploiting the nonlinearities induced in both resonators, we use the transmon qubit to track the parity of the storage cavity state. For simplicity, we will refer to the storage cavity as ``the cavity" henceforth.

The qubit and cavity are in the strong dispersive coupling regime, which can be described by the Hamiltonian:
\begin{equation*}
H/\hbar=\omega_q\ket{e}\bra{e}+(\omega_s-\chi_{qs}\ket{e}\bra{e})a^\dagger a
\label{eq:Hamiltonian}
\end{equation*}
where $a$ and $a^\dagger$ are the annihilation and creation operators respectively, $\ket{e}$ is the excited state of the qubit, and $\chi_{qs}/2\pi=1.789$~MHz is the qubit state dependent frequency shift of the cavity.  The readout cavity has been neglected because it remains in the ground state while the system actually evolves. The interaction between the qubit and the cavity entangles qubit and photon. In the rotating frame of the cavity, Fock states associated with the qubit in the excited state acquire a phase $\Phi=a^\dagger a\chi_{qs}t$ proportional to their photon number~\cite{Schuster}.  By waiting for $t=\pi/\chi_{qs}$, one can realize a controlled-phase gate $C_\pi=I\bigotimes\ket{g}\bra{g}+e^{i\pi a^\dagger a}\bigotimes\ket{e}\bra{e}$, adding a $\pi$ phase shift per photon on the cavity state conditioned on the qubit state~\cite{Vlastakis}. Therefore, $C_\pi$ can be inserted between two $\pi/2$ pulses on the qubit in a Ramsey-type measurement to map the photon parity of any cavity state onto the qubit (black dashed line enclosure in Fig.~1b). The result of a qubit measurement after the second $\pi/2$ pulse together with prior knowledge of the initial qubit state indicates whether the number of photons in the cavity is even or odd, but reveals nothing about the actual value of the photon number.

The creation of cat states is a natural consequence of a parity measurement on a coherent state $\ket{\alpha}$ since the phase cat states $\ket{\alpha}\pm\ket{-\alpha}$ are eigenstates of the parity operator $ e^{i\pi a^\dagger a}$~\cite{Brune}. After displacing the cavity vacuum by $\alpha$ with the qubit initially at $\ket{g}$, we use the parity protocol to take $(\ket{\alpha, g}+\ket{\alpha, e})/\sqrt{2}$ after the first $\pi/2$ pulse to $[(\ket{\alpha}-\ket{-\alpha})\ket{g}$+$(\ket{\alpha}+\ket{-\alpha})\ket{e}]/2$ after the second pulse, at which point the parity of the cavity state is entangled with the state of the qubit. Detection of the qubit state using the readout cavity then projects the storage cavity onto one of the two cat states. To confirm the non-classical properties of these states, we perform Wigner tomography of the cavity after a single parity measurement for an initial coherent state of displacement $|\alpha|=2$ ($\bar{n}=4$).  Post-selecting on the ground or excited qubit states to obtain the odd or even cats (Figs.~2a and 2b), respectively, we see the interference patterns that are the signature of quantum behavior. The overlap between the measured Wigner function and that of an ideal cat gives a fidelity of $F=83\%$. Figure~2c shows Wigner function without post selection (tracing over qubit states). Fringes in the Wigner function completely disappear as expected and we obtain the statistical mixture of even and odd states. The high contrast between even and odd cat states is a central requirement in implementing a recently proposed QEC scheme~\cite{Leghtas2}, where these form the code and error spaces respectively.

As the loss of a single photon changes the parity of a cat state, monitoring parity repeatedly in real time allows us to track photon jumps of our cavity. Here we note that to interpret the result of a single parity measurement we must know the state of the qubit prior to the first $\pi/2$ pulse.  In other words, it is the correlation of the qubit states before and after the parity measurement (an oscillating pattern between $\ket{g}$ and $\ket{e}$ vs. a constant pattern remaining in either $\ket{g}$ or $\ket{e}$) that reveals the photon state parity. For the following data we have chosen $R_{\hat{y},-\frac{\pi}{2}}$ as the second qubit pulse, instead of $R_{\hat{y},\frac{\pi}{2}}$, in order to maintain a constant pattern when the cavity is in the even parity state. Aside from reversing which pattern we assign to be even and odd, this change makes no difference. Figures~3b-e show typical 400 $\mu$s single-shot traces. The initial displacement is $|\alpha|=1.0$ and the repetition interval of the parity measurements is 1~$\mu$s, much smaller than the average photon lifetime $\tau_0=55~\mu$s obtained from a free time evolution measurement of the parity of a coherent state (see Supplementary Material A). We observe a range of photon jump statistics, from quiet traces that last for hundreds of microseconds with no apparent changes in parity, to those that have as many as five jumps. The clear dichotomy between the patterns in our traces indicates that although the measurements are susceptible to qubit decoherence, as evidenced by intermittent brief changes in measurement correlations and excitations to higher qubit states, they nonetheless exhibit a strong sensitivity to single photon jump events. 

\begin{figure}[t]
\includegraphics{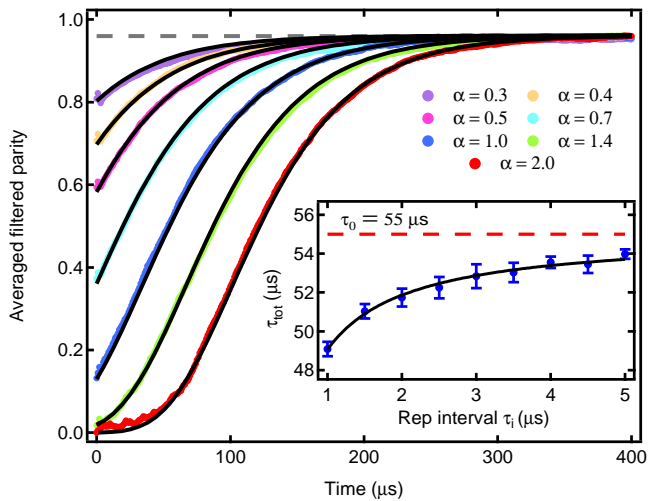}
\caption {Ensemble averaged dynamics of the parity estimator without distinguishing the initial parity of the created cats. The repetition time of the parity measurement $\tau_i=1~\mu$s. The dashed horizontal line represents the expected saturation of the parity due to a background photon number $n_{th}=0.02$ in the cavity from an independent measurement (see Supplementary Material A). Solid lines are fitted theoretical curves for coherent states with $n_{th}$, but with the time constant (a global fitting gives $\tau_{\rm tot}=49~\mu$s) as the only free parameter (see Supplementary Material B). This $\tau_{\rm tot}$ value closely matches the expected lifetime $\tau_0$, obtained from a free evolution measurement. Inset: extracted time constants as a function of different parity measurement repetition intervals. The decay time $\tau_{\rm tot}$ is modelled as $1/\tau_{\rm tot}=1/\tau_0+P_{D}/\tau_i$, where $P_{D}$ is the probability of inducing an extra parity change. A fit (solid line) gives $P_{D}$=$2\times 10^{-3}$, indicating a 99.8\% QND of each parity measurement. Moreover, as each averaged parity decay saturates at the expected value (1-2$n_{th}$) indicates that repeatedly measuring the cavity does not raise its thermal population, further confirming how QND our measurements are.}
\end{figure}

When analyzing these single-shot traces, in order to mitigate the effects due to qubit decoherence, excitation to qubit states higher than $\ket{e}$ (denoted as $\ket{f}$), and other imperfections in the qubit readout in extracting the parity, we have applied a quantum filter that best estimates the photon state parity (details described in Supplementary Material B). We note that the quantum filter output depends on the entire previous parity trajectory. Figures~3b-e show traces with the parity estimator calculated from the quantum filter, in red. The parity estimator is clearly much less sensitive to qubit decoherence and $\ket{f}$ states. Although our single parity readout fidelity is 80\%, due to the smoothing effect of the quantum filter, we actually can achieve nearly unity detection sensitivity of single photon jumps. However, given one jump, the probability to have a second jump within the response time of the filter ($\sim2~\mu$s) is 4\% for $\bar{n}=1$ (or 15\% for $\bar{n}=4$), which limits our overall detection sensitivity over an entire trajectory (see Supplementary Material C).

The repeated parity measurements shown above constitute just a single point, the origin, in the Wigner functions of the even and odd cats (Figs.~2a and 2b). The loss of a single photon flips not just the parity, but the sign of the entire fringe pattern as well. Thus, crucially, a parity measurement acquires no information about the phase of the cat states. Consequently, one could encode quantum information onto the computational bases $\ket{0}_L=\mathcal{N}(\ket{\alpha}+\ket{-\alpha})$ and $\ket{1}_L=\mathcal{N}(\ket{i\alpha}+\ket{-i\alpha})$ with $\mathcal{N}=1/\sqrt{2(1+e^{-2|\alpha|^2})}$, and any subsequent parity measurements would make no distinction between the two. It is this capability of extracting error syndromes without perturbing the encoded information that is so essential to QEC.

The degree to which the measurements are QND can be determined by examining the decay rate for the parity of a coherent state with different measurement cadences. We extract the total decay rate of the parity ($\tau_{\rm tot}$), from the ensemble averaged parity dynamics obtained with the quantum filter (Fig.~4). This total decay rate is well modelled by the parallel combination of the free decay time ($\tau_0 = 55~\mu$s) plus a constant demolition probability $P_D = 0.002$ per measurement interval $\tau_i$, as shown by the fit in the inset of Fig.~4. In other words, a single parity measurement is 99.8\% QND. 

Further improvements of this measurement technique will be required to realize a truly robust error-corrected quantum memory. The probability of missing a photon jump, due to the finite measurement rate per cavity lifetime, would be greatly reduced if combined with longer lived cavities~\cite{Reagor}. In addition, the current approach is not yet fault-tolerant, since relaxation (finite qubit $T_1$) of the ancilla induces phase errors in the cat states. Improving these lifetimes and further characterizing these types of error processes are important next steps. Nonetheless, we estimate that when combined with an optimized measurement strategy, the current level of performance could already allow an extension of the lifetime ($>\tau_{\rm tot}/\bar{n}$) for an encoded cat state by over a factor of two (see Supplementary Material D).

In summary, we have demonstrated the real-time tracking of jumps in the photon number parity in circuit QED. Significantly, this quantity differs from previous observations of quantum jumps between energy levels. Rather, it projects the system into a degenerate subspace, and can therefore serve as an error syndrome for QEC. We show that the parity measurement is highly QND, and has a high fidelity and cadence compared to the cavity lifetime. These performances represent a significant advance in the measurement capabilities necessary for further progress in quantum information.

\vspace{0.2in}
\textbf{Acknowledgements}
We thank L. Jiang and S. M. Girvin for helpful discussions. Facilities use was supported by the Yale Institute for Nanoscience and Quantum Engineering (YINQE) and the NSF MRSEC DMR 1119826. This research was supported in part by the Office of the Director of National Intelligence (ODNI), Intelligence Advanced Research Projects Activity (IARPA), through the Army Research Office (W911NF-09-1-0369) and in part by the U.S. Army Research Office (W911NF-09-1-0514). All statements of fact, opinion or conclusions contained herein are those of the authors and should not be construed as representing the official views or policies of IARPA, the ODNI, or the U.S. Government. MM acknowledges partial support from the Agence National de Recherche under the project EPOQ2, ANR-09-JCJC-0070.

\vspace{0.2in}
\textbf{Author contributions}
LS and AP performed the experiment and analyzed the data. ZL and MM provided theoretical support. BV and GK provided further experimental contributions. KS, AN, MH, and SS contributed to the double-pumped Josephson bifurcation amplifier under the supervision of MHD. JB and LF fabricated the device. RJS designed and supervised the project. LS, AP, LF, and RJS wrote the manuscript with feedback from all authors.

\vspace{0.2in}
\textbf{Correspondence}  Correspondence and requests for materials should be addressed to LS (email: luyansun@mail.tsinghua.edu.cn) or RJS (email: robert.schoelkopf@yale.edu)

\end{document}


\title{Supplementary Material for ``Tracking Photon Jumps with Repeated Quantum Non-Demolition Parity Measurements"}
\author{L.~Sun\footnote{current address: Center for Quantum Information, Institute for Interdisciplinary Information Sciences, Tsinghua University, Beijing, P. R. China}}
\affiliation{Departments of Applied Physics and Physics, Yale University, New Haven, CT 06511, USA}
\author{A.~Petrenko}
\affiliation{Departments of Applied Physics and Physics, Yale University, New Haven, CT 06511, USA}
\author{Z.~Leghtas}
\affiliation{Departments of Applied Physics and Physics, Yale University, New Haven, CT 06511, USA}
\author{B.~Vlastakis}
\affiliation{Departments of Applied Physics and Physics, Yale University, New Haven, CT 06511, USA}
\author{G.~Kirchmair\footnote{current address: Institut f\"{u}r Experimentalphysik, Universit\"{a}t Innsbruck, Technikerstra{\ss}e 25, A-6020 Innsbruck, Austria;
Institut f\"{u}r Quantenoptik und Quanteninformation, \"{O}sterreichische Akademie der Wissenschaften, Otto-Hittmair-Platz 1, A-6020 Innsbruck, Austria}}
\affiliation{Departments of Applied Physics and Physics, Yale University, New Haven, CT 06511, USA}
\author{K.~M.~Sliwa}
\affiliation{Departments of Applied Physics and Physics, Yale University, New Haven, CT 06511, USA}
\author{A.~Narla}
\affiliation{Departments of Applied Physics and Physics, Yale University, New Haven, CT 06511, USA}
\author{M.~Hatridge}
\affiliation{Departments of Applied Physics and Physics, Yale University, New Haven, CT 06511, USA}
\author{S.~Shankar}
\affiliation{Departments of Applied Physics and Physics, Yale University, New Haven, CT 06511, USA}
\author{J.~Blumoff}
\affiliation{Departments of Applied Physics and Physics, Yale University, New Haven, CT 06511, USA}
\author{L.~Frunzio}
\affiliation{Departments of Applied Physics and Physics, Yale University, New Haven, CT 06511, USA}
\author{M.~Mirrahimi}
\affiliation{Departments of Applied Physics and Physics, Yale University, New Haven, CT 06511, USA}
\affiliation{INRIA Paris-Rocquencourt, Domaine de Voluceau, B.P.~105, 78153 Le Chesnay Cedex, France}
\author{M.~H.~Devoret}
\affiliation{Departments of Applied Physics and Physics, Yale University, New Haven, CT 06511, USA}
\author{R.~J.~Schoelkopf}
\affiliation{Departments of Applied Physics and Physics, Yale University, New Haven, CT 06511, USA}

\maketitle

\subsection{Experiment setup, device parameters, and readout properties}
Our measurements are performed in a cryogen-free dilution refrigerator with a base temperature of about 10~mK. Figure~S1 shows the schematic of the measurement setup. A Josephson bifurcation amplifier (JBA)~\cite{Siddiqi,Vijay2} operating in a double-pumped mode~\cite{Kamal,Murch,Sliwa} is used as the first stage of amplification between the readout cavity output and the high electron mobility transistor (HEMT), allowing for a high-fidelity single-shot dispersive readout of the qubit state. We typically operate the JBA in the saturated regime with about 20 readout photons for a better signal-to-noise ratio.

The transmon qubit is fabricated on a $c$-plane sapphire (Al$_2$O$_3$) substrate with a double-angle evaporation of aluminum after a single electron-beam lithography step. The qubit has a transition frequency $\omega_q/2\pi=5.938$ GHz with an anharmonicity $\alpha_q/2\pi=(\omega_{ge}-\omega_{ef})/2\pi$=240 MHz, an energy relaxation time $T_1=8~\mu$s and a Ramsey time $T_2^*=5~\mu$s. Even at the lowest base temperature, the qubit is measured to have about 86\% ground state $\ket{g}$, 11\% excited state $\ket{e}$, and 3\% of states higher than $\ket{e}$, denoted as $\ket{f}$. These excitations of the qubit could come from stray infrared photons leaking into the cavity, although the exact source remains unknown.

The qubit serves as an ancilla and provides the necessary non-linearity for the manipulation of coherent states in the storage cavity. Both the storage and readout cavities are made of aluminum alloy 6061. The state dependent frequency shifts between the qubit and the storage and readout cavities are $\chi_{qs}/2\pi=1.789$ MHz and $\chi_{qr}/2\pi=0.930$ MHz respectively. For simplicity, we will refer to the storage cavity as ``the cavity" henceforth. The inset of Fig.~S2 shows the so-called number splitting peaks of the qubit due to different photon numbers in the cavity, which is displaced with a 10~ns square pulse right before the spectroscopy measurement. A second order polynomial fit $\chi(N)=-\chi_{qs}N+\chi_{qs}'N^2$ gives a non-linear correction to the dispersive shift~\cite{Vlastakis} $\chi_{qs}'/2\pi=1.9\pm0.1$~kHz which is small enough to be neglected in the cavity dynamics. Figure~S2 shows the probability of the first seven Fock states $n=0,1,2,...6$ as a function of displacement amplitude $|\alpha|$ in excellent agreement with a Poisson distribution, indicating a good control of the coherent state in the cavity. We scale the x-axis from the voltage amplitude of the displacement pulse applied from an arbitrary waveform generator and use this scaling as a calibration. There is a small residual amplitude for the $n=1$ peak even with no displacement (point near origin), allowing us to infer that there is a background photon population $n_{th}=0.02$ in the cavity. The lifetime of the cavity is characterized by measuring a free parity evolution of a coherent state as shown in Fig.~S3, which is nearly identical to Fig.~4 in the main text. A global fitting gives a time constant $\tau_0=55~\mu$s.

\begin{figure}
\includegraphics{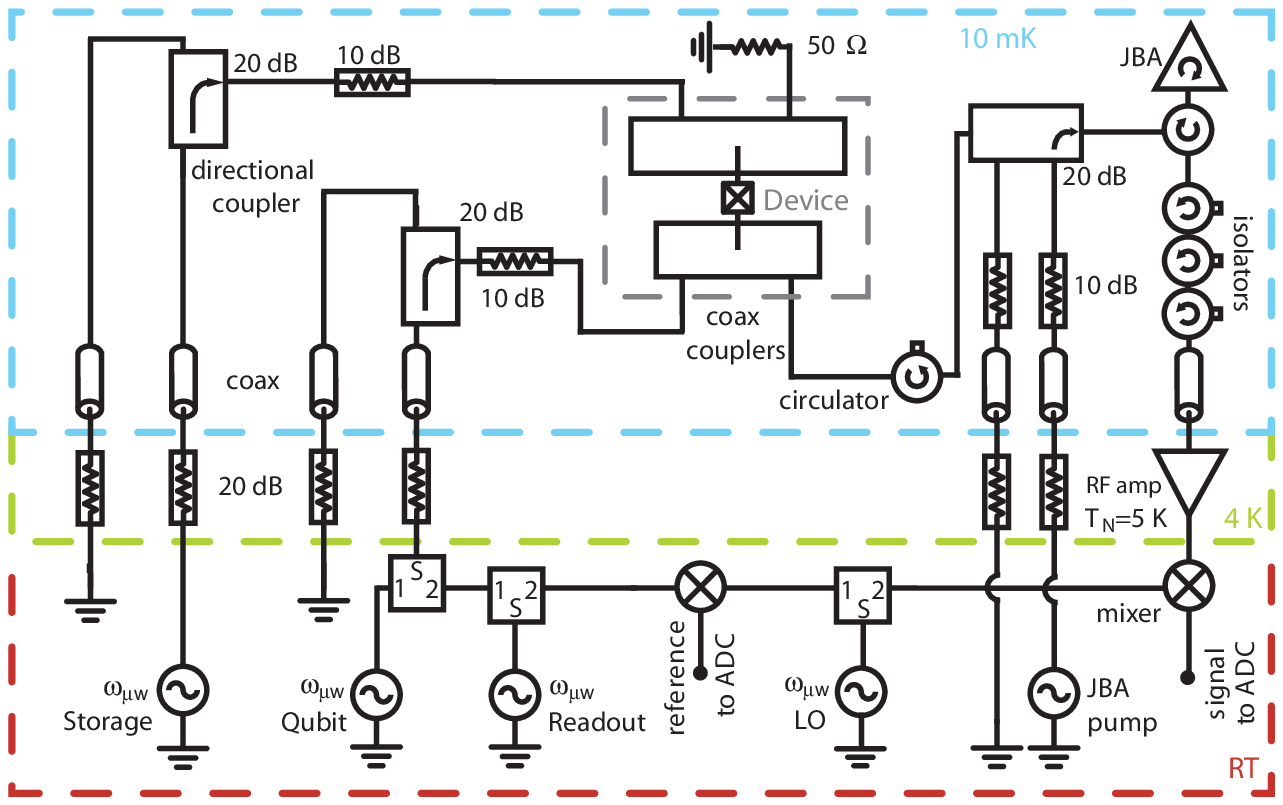}
\caption {Schematic of the measurement setup. We use two separate lines to drive the readout and the storage cavity. Qubit state manipulations are realized through the readout cavity input line. The readout cavity output signal is first amplified by a JBA operating in a double-pumped mode, and the reflected signal then goes through three isolators in series before being further amplified by a HEMT at 4 K. The amplified signal is finally down-converted to 50~MHz and then digitized by a fast 1 GS data-acquisition card.}
\end{figure}

\begin{figure}
\includegraphics{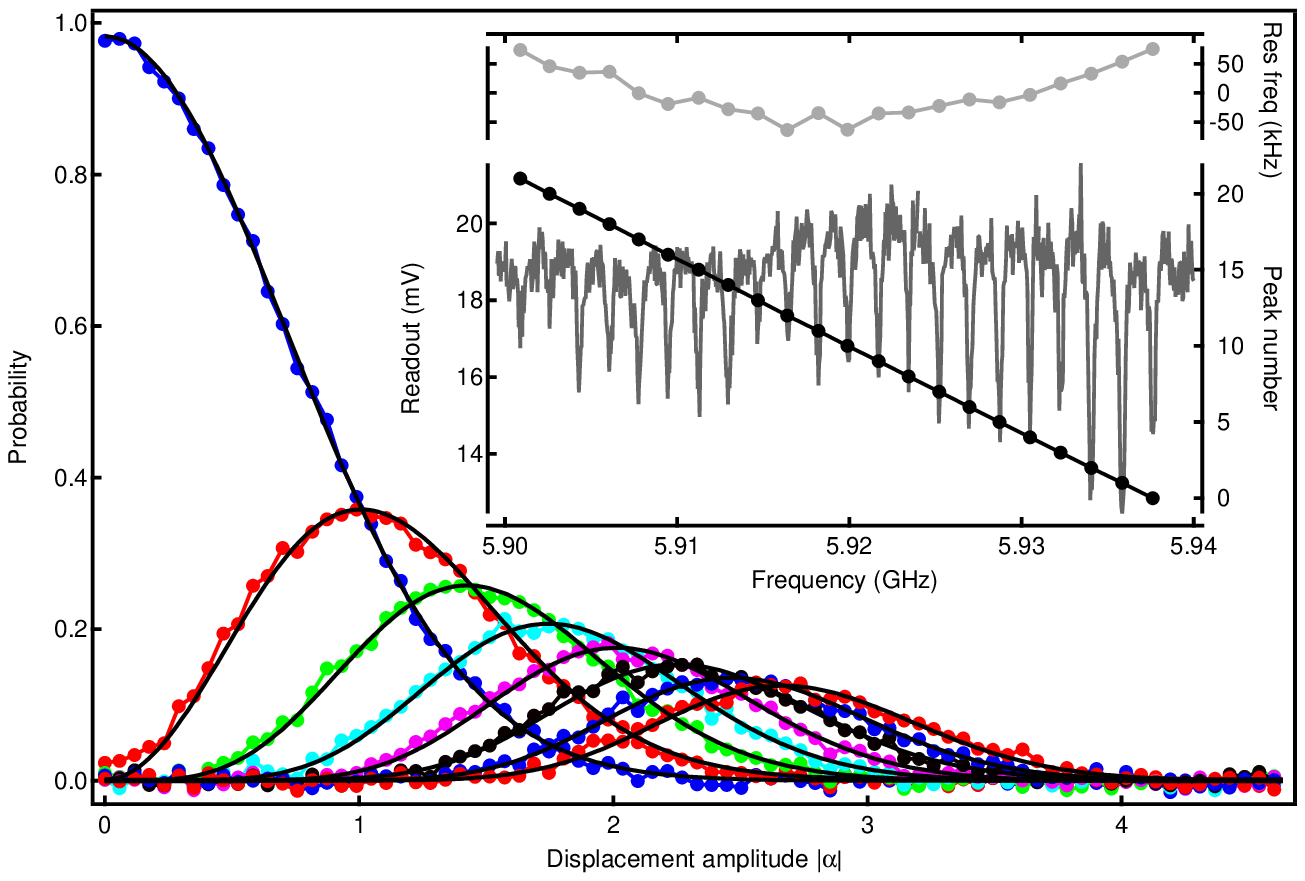}
\caption {Poisson distribution of photon numbers in the cavity. Dotted color lines are data for the first seven Fock states $n=0,1,2,...6$ as a function of displacement amplitude $|\alpha|$. The measurements are performed with a selective $\pi$ pulse on each number splitting peak and the resulting signal amplitude should be proportional to the corresponding number population. These oscillation amplitudes have been normalized to probabilities such that the sum of the amplitudes corresponding to $n=0$ and $n=1$ equals to unity. Dashed lines are theoretical curves with a Poisson distribution $\mathbb{P}(|\alpha|)=|\alpha|^{2n}e^{-|\alpha|^2}/n!$ where the x-axis has had a single scale factor adjusted to fit all these probabilities. The excellent agreement indicates a good control of the coherent state in the cavity. Based on the probability of $n=1$ at $|\alpha|=0$, we find a background photon population $n_{th}=0.02$ in the cavity. Inset: spectroscopy (left axis) of the number splitting peaks of the qubit when populating different photon numbers in the cavity. Top panel shows the difference between peak positions and a linear fit. The curvature necessitates a second order polynomial fit resulting a linear dispersive shift $\chi_{qs}/2\pi=1.789\pm0.002$~MHz and a non-linear dispersive shift $\chi_{qs}'/2\pi=1.9\pm0.1$~kHz.}
\end{figure}

We have adjusted the phase between the JBA readout signal and the pump such that $\ket{g}$, $\ket{e}$, and $\ket{f}$ states can be distinguished with optimal contrast.  Figure~S5a shows the histogram of the qubit readout for the parity protocol used in repeated single-shot traces in Fig.~3 in the main text. The histogram is clearly trimodal. Thresholds between $\ket{g}$ and $\ket{e}$, and between $\ket{e}$ and $\ket{f}$ states have been chosen to digitize the readout signal to $+1, -1$, and $0$ for $\ket{g}$, $\ket{e}$, and $\ket{f}$ state respectively. We assign a zero to the $\ket{f}$ states to indicate a ``failed" measurement with no useful information about the parity. These $\ket{f}$ states can be fixed with a field programmable gate array applying proper pulses to drive the qubit back to either $\ket{g}$ or $\ket{e}$ in real time. Figure~S5b shows the basic qubit readout properties with the cavity left in vacuum. The $\ket{g}$ state is prepared through a post-selection of an initial qubit measurement, while $\ket{e}$ and $\ket{f}$ are prepared by properly pulsing the selected $\ket{g}$ state. The loss of fidelity predominantly comes from the $T_1$ process during both the waiting time of the initialization measurement (300~ns) and the qubit readout time (340~ns).

\begin{figure}
\includegraphics{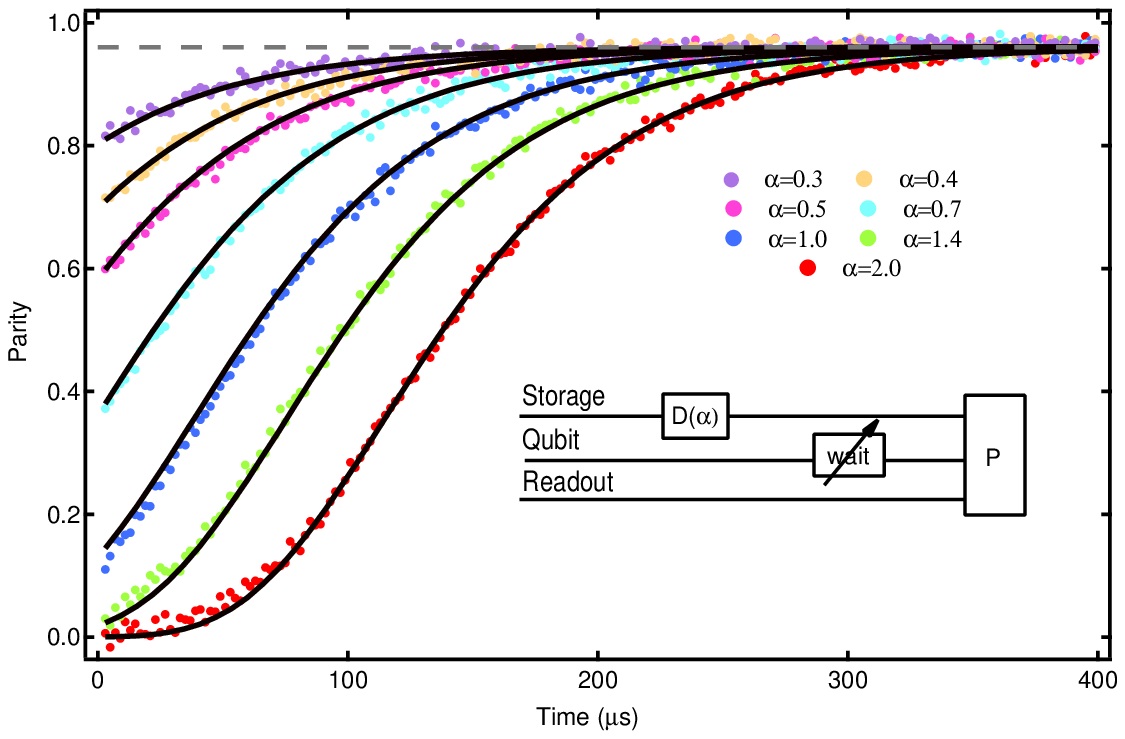}
\caption{Ensemble averaged free parity evolution of a coherent state. The measurement protocol is shown in the inset. The single parity measurement gives a readout voltage that has been converted to parity through thresholding. All measured evolution curves saturate at the same value in the long time limit. This saturation level has been forced to 0.96 (due to $n_{th}=0.02$), represented by the dashed Horizontal line. The solid lines are global fits, giving a time constant $\tau_0=55~\mu$s.}
\end{figure}

To perform a good parity measurement, the $\pi/2$ pulses $R_{\hat{y},\pm\frac{\pi}{2}}$ should equally cover as many number splitting peaks as possible without significantly exciting the $\ket{f}$ state. We choose a Gaussian envelope pulse truncated to $4\sigma=8~ns$ ($\sigma_f=80$~MHz) for a good compromise. Figure~S4 shows the effectiveness of those $R_{\hat{y},\pm\frac{\pi}{2}}$ pulses as a function of $\bar{n}$ in the cavity. The curvature for $\bar{n}>4$ is due to the finite bandwidth of those pulses in the frequency domain.

\begin{figure}
\includegraphics{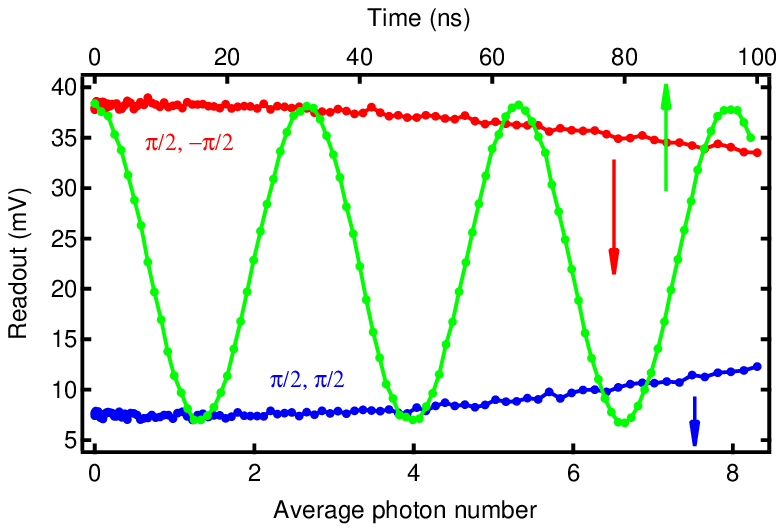}
\caption {Effectiveness of $R_{\hat{y},\pm\frac{\pi}{2}}$ pulse. Blue and red data (bottom axis) are ensemble averaged qubit readout after consecutively (with no wait time) applying ($R_{\hat{y},\frac{\pi}{2}}$, $R_{\hat{y},\frac{\pi}{2}}$) and ($R_{\hat{y},\frac{\pi}{2}}$, $R_{\hat{y},-\frac{\pi}{2}}$) respectively as a function of different $\bar{n}$ introduced into the cavity. The curvature for $\bar{n}>4$ comes from the finite bandwidth of the pulses in the frequency domain. Green curve (top axis) is a time Rabi trace for an amplitude comparison with no initial cavity displacement.}
\end{figure}

We emphasize that it is the correlation $C_t$ of the qubit states before and after the parity measurement that reveals the photon state parity. Figure~S5c shows the parity readout properties of our system. The loss of fidelity of the parity measurement mainly comes from qubit decoherence process during the parity measurement. Conditional probabilities $\mathbb{P}(+1|even)$, $\mathbb{P}(+1|odd)$, $\mathbb{P}(-1|even)$, $\mathbb{P}(-1|odd)$, $\mathbb{P}(0|even)$, and $\mathbb{P}(0|odd)$ are time-independent probabilities which have a positive, negative, and zero correlation between the digitized qubit readouts before and after a parity measurement for a given even or odd state. However, a pure even or odd state cannot be prepared easily in our system due to the finite thermal population of the cavity, which is small but can still introduce systematic errors. We determine $\mathbb{P}(\pm1,0|even/odd)$ by post-selecting the cases with five consecutive identical parity results, which give a good confidence of the photon state parity, and then performing a histogram on the sixth parity measurement. 

\begin{figure}
\includegraphics{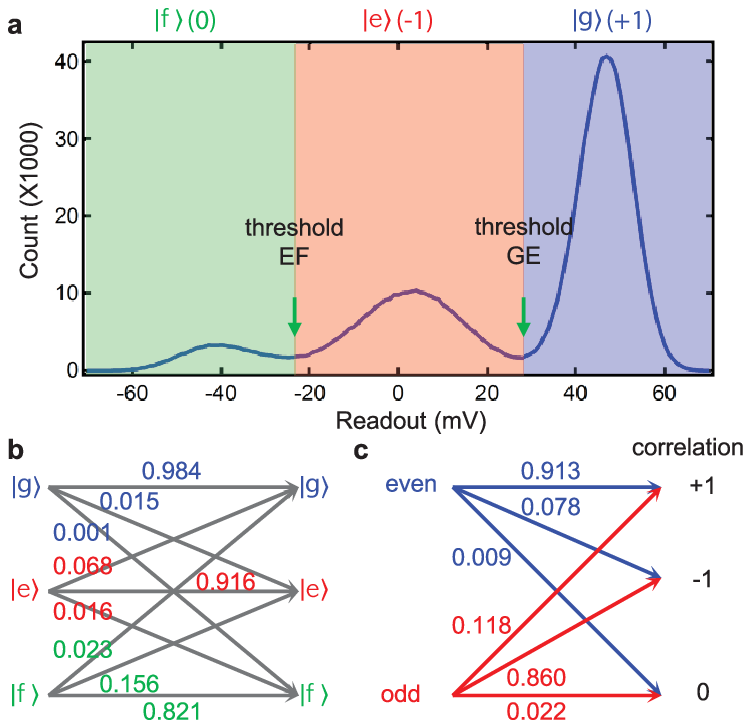}
\caption {(a) Histogram of qubit readout for the parity protocol used in repeated single-shot traces in Fig.~3 in the main text. The phase between the JBA readout and the pump has been adjusted such that $\ket{g}$, $\ket{e}$, and $\ket{f}$ states can be distinguished with optimal spacings. Thresholds between $\ket{g}$ and $\ket{e}$, and between $\ket{e}$ and $\ket{f}$ have been chosen to digitize the readout signal to $+1, -1$, and $0$ for $\ket{g}$, $\ket{e}$, and $\ket{f}$ respectively. Note that we assign a zero to the $\ket{f}$ states to indicate a ``failed" measurement with no useful information about the parity. (b) Qubit readout properties for an initial qubit state at $\ket{g}$, $\ket{e}$, and $\ket{f}$ state respectively. (c) Parity readout property for a given even and odd parity state. $\mathbb{P}(\pm1,0|even/odd)$ are determined by post-selecting the cases with five consecutive identical parity results, which give a good confidence of the photon state parity, and then performing a histogram on the sixth parity measurement.}
\end{figure}

\subsection{Quantum filter and correlated data}
\begin{figure}
\includegraphics{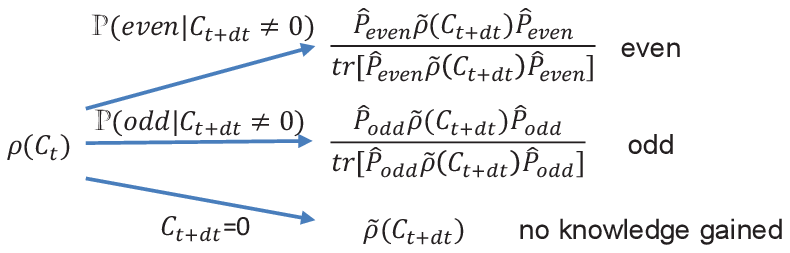}
\caption {Schematic of the quantum filter. At time $t$, the density matrix of the photon state is $\rho(C_{t})$, which depends on all previous correlations. At $t+dt$, only considering the decoherence of the cavity, the expected density matrix from free evolution becomes $\tilde{\rho}(C_{t+dt})$. The additional information $C_{t+dt}$ acquired from the parity measurement at $t+dt$ changes the knowledge of the parity of the photon state according to Eq.~S\ref{eq:QuantumFilter1}.}
\end{figure}

In order to mitigate the effects due to qubit decoherence, $\ket{f}$ state of the qubit (an undesirable state that obscures the parity measurement), and other imperfections in the qubit readout in extracting the parity, we have applied a quantum filter that best estimates the photon state parity. We note that the quantum filter depends on the measured trajectory, that is on the past history of measurement results. Figure~S6 shows the schematic of the quantum filter. This quantum filter at each point in time is realized in two steps: first, a new density matrix $\tilde{\rho}(C_{t+dt})$ is calculated from the best estimation $\rho(C_t)$ at the previous point based only on the decoherence of the cavity; second, the density matrix $\tilde{\rho}(C_{t+dt})$ gets updated as the best estimation $\rho(C_{t+dt})$ according to Bayes law based on the newly acquired knowledge from the current parity measurement. This best estimated density matrix $\rho(C_{t+dt})$ is then used as the input for the next iteration. We have truncated the dimension of the density matrix to $N=5\bar{n}$, which is large enough to cover all relevant number states. To initialize the density matrix after a displacement $D(\alpha)$, we have set ${\rho}(t=0)=(1-n_{th})D(\alpha)\ket{0}\bra{0}D^\dagger(\alpha)+n_{th}D(\alpha)\ket{1}\bra{1}D^\dagger(\alpha)$, taking into account the background photon population in the limit $n_{th}\ll1$.

At time $t$, the density matrix of the photon state is $\rho(C_t)$, which depends on all previous correlations. At $t+dt$, only considering the decoherence of the cavity, the expected density matrix from free evolution becomes $\tilde{\rho}(C_{t+dt})=M_{down}\rho(C_{t}) M_{down}^\dagger+M_{up}\rho(C_{t}) M_{up}^\dagger+M_{no}\rho(C_{t}) M_{no}^\dagger$, where $M_{down}=\sqrt{\kappa_{down}dt}a$, $M_{up}=\sqrt{\kappa_{up}dt}a^\dagger$, and $M_{no}=I-(M_{down}^\dagger M_{down}+M_{up}^\dagger M_{up})/2$ are the Kraus operators for photon loss, absorption of thermal photons, and no jump events respectively. We have $\kappa_{down}=(n_{th}+1)\kappa$ and $\kappa_{up}=n_{th}\kappa$, and $\kappa=1/\tau_{\rm tot}$ is the energy decay rate in the cavity under repeated parity measurements. The additional information $C_{t+dt}$ acquired from the parity measurement at $t+dt$ changes the quantum state according to:
\begin{equation}
\rho(C_{t+dt})=
\begin{cases}
\mathbb{P}(even|C_{t+dt})\frac{\hat{P}_{even}\tilde{\rho}(C_{t+dt})\hat{P}_{even}}{tr(\hat{P}_{even}\tilde{\rho}(C_{t+dt})\hat{P}_{even})}+\mathbb{P}(odd|C_{t+dt})\frac{\hat{P}_{odd}\tilde{\rho}(C_{t+dt})\hat{P}_{odd}}{tr(\hat{P}_{odd}\tilde{\rho}(C_{t+dt})\hat{P}_{odd})}, & \text{if $C_{t+dt}\neq 0$,}\\
    \tilde{\rho}(C_{t+dt}) & \text{if $C_{t+dt}=0.$}
\end{cases}
\label{eq:QuantumFilter1}
\end{equation}
where $\hat{P}_{even}$ and $\hat{P}_{odd}$ are the projectors onto the even and odd manifolds, $\hat{P}=\hat{P}_{even}-\hat{P}_{odd}=e^{i\pi a^\dagger a}$ is the parity operator, $\mathbb{P}(even|C_{t+dt})$ and $\mathbb{P}(odd|C_{t+dt})$ are the probabilities of being in the even and odd parity respectively for a measured $C_{t+dt}$. To simplify the quantum filter, we assume that the event of the qubit jumping to the $\ket{f}$ states is independent of the cavity parity being even or odd. Hence, if the measured correlation is zero, the density matrix of the photon state is assigned to the expected one from free evolution. Based on Bayes law, Eq.~S\ref{eq:QuantumFilter1} becomes:
\begin{equation}
\rho(C_{t+dt})=
\begin{cases} \frac{\mathbb{P}(C_{t+dt}|even)\hat{P}_{even}\tilde{\rho}(C_{t+dt})\hat{P}_{even}+\mathbb{P}(C_{t+dt}|odd)\hat{P}_{odd}\tilde{\rho}(C_{t+dt})\hat{P}_{odd}}{\mathbb{P}(C_{t+dt})}, & \text{if $C_{t+dt}\neq 0$,}\\
    \tilde{\rho}(C_{t+dt}) & \text{if $C_{t+dt}=0.$}
\end{cases}
\label{eq:QuantumFilter2}
\end{equation}
where $\mathbb{P}(C_{t+dt})=\mathbb{P}(C_{t+dt}|even)tr[\hat{P}_{even}\tilde{\rho}(C_{t+dt})\hat{P}_{even}]+\mathbb{P}(C_{t+dt}|odd)tr[\hat{P}_{odd}\tilde{\rho}(C_{t+dt})\hat{P}_{odd}]$. The best parity estimation of the photon state is then: 
\begin{equation}
\rm{P}\mathnormal{(t+dt)=tr[\rho(C_{t+dt})\hat{P}]}
\label{eq:QuantumFilter3}
\end{equation}
This formula has been used extensively in the main text to estimate the parity of the photon state. 

In order to make a comparison with the best parity estimation based on the above quantum filter, we also directly correlate the neighboring parity measurements without any further processing. For zero correlation cases, since no information of the photon state parity is acquired, the best knowledge of parity at those points is just the last measured non-zero correlation. We assume the repeated parity measurement is a Markovian process. The ensemble averaged parity dynamics obtained from the correlation under a repeated parity monitoring is then simply:
\begin{align}
<C_{cor}(t)>=\mathbb{P}(+1,t)-\mathbb{P}(-1,t)+\mathbb{P}(0,t)\frac{\mathbb{P}(+1,t)-\mathbb{P}(-1,t)}{\mathbb{P}(+1,t)+\mathbb{P}(-1,t)}
\label{eq:Correlated_results}
\end{align}
where $\mathbb{P}(+1,t)$, $\mathbb{P}(-1,t)$, and $\mathbb{P}(0,t)$ are the probability of measuring positive, negative, and zero correlations at time $t$. The third term comes from the fact that the cases with zero correlation is assigned to previously measured non-zero correlation +1 or -1 whose probability is $\mathbb{P}(+1,t-\Delta t)$ and $\mathbb{P}(-1,t-\Delta t)$. For small $\Delta t$, $\mathbb{P}(\pm1,t-\Delta t) \approx \mathbb{P}(\pm1,t)$.

The probabilities $\mathbb{P}(+1,t)$, $\mathbb{P}(-1,t)$, and $\mathbb{P}(0,t)$ depend on both the measured parity readout property $\mathbb{P}(\pm1,0|even/odd)$ and the even and odd parity evolution $\rm{P}_e(\mathnormal t)$ and $\rm{P}_o(\mathnormal t)$ of the photon state:
\begin{equation}
\begin{array}{l}
\displaystyle \mathbb{P}(+1,t)=\mathbb{P}(+1|even)\rm{P}_e(\mathnormal t)+\mathbb{P}(+1|odd)\rm{P}_o(\mathnormal t)\\
\displaystyle \mathbb{P}(-1,t)=\mathbb{P}(-1|even)\rm{P}_e(\mathnormal t)+\mathbb{P}(-1|odd)\rm{P}_o(\mathnormal t)\\
\displaystyle \mathbb{P}(0,t)=\mathbb{P}(0|even)\rm{P}_e(\mathnormal t)+\mathbb{P}(0|odd)\rm{P}_o(\mathnormal t)
\end{array}
\label{eq:P_plus_minus_zero}
\end{equation}
where $\rm{P}_e(\mathnormal t)=\mathnormal {(e^{-2|\alpha|^2 e^{-\kappa t}}+1)/2}$ and $\rm{P}_o(\mathnormal t)=\mathnormal {(1-e^{-2|\alpha|^2 e^{-\kappa t}})/2}$.

With all the parameters in Eq.~S\ref{eq:Correlated_results} known, $<C_{cor}(t)>$ can then be predicted. The agreement with the measured data is excellent as shown in  Fig.~S7. This data set is the same as that shown in Fig.~4 in the main text. Equation~S\ref{eq:Correlated_results} even accurately predicts the offset in the averaged parity at $t=0$ which comes from the asymmetric parity readout fidelities between the even and odd states. The fact that the saturated parity value in the long time limit in Fig.~S7 is much lower than that in Fig.~4 in the main text mainly comes from the qubit decoherence and the imperfections in the qubit readout. This large difference is additional proof of the effectiveness of the quantum filter. 

For a coherent state in a thermal bath, its parity dynamics is simply~\cite{Haroche}:
\begin{equation}
\rm{P}=\mathnormal{\frac{1}{1+2n_{th}}e^{-2|\alpha|^2 e^{-\kappa t}/(1+2n_{th})}},
\label{eq:Parity_coherent_with_nBG}
\end{equation}
which has been used to fit Fig.~4 in the main text.

\begin{figure}
\includegraphics{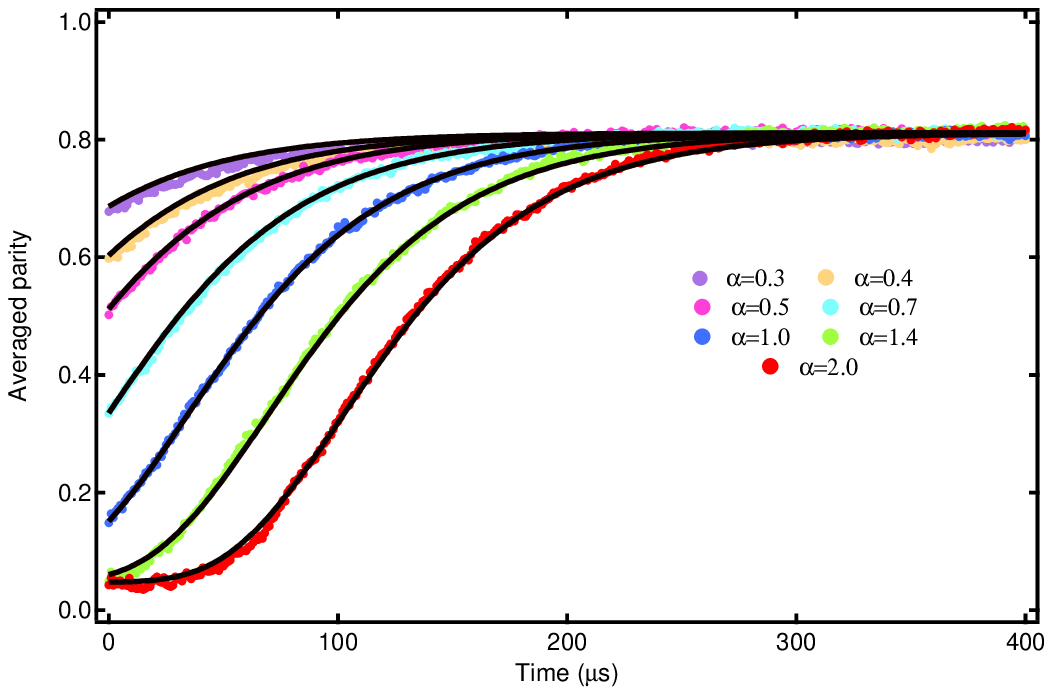}
\caption{Ensemble averaged parity dynamics obtained directly from the correlation of qubit states between neighboring parity measurements under a repeated parity monitoring. The data set is the same as that shown in Fig.~4 in the main text. Solid lines are predictions based on Eq.~S\ref{eq:Correlated_results}, in excellent agreement with the measured data. The offset of the averaged parity at $t=0$ comes from the asymmetric parity readout fidelities between the even and odd states. The fact that the saturated parity value in the long time limit is much lower than that in Fig.~4 in the main text mainly comes from the qubit decoherence and the imperfections in the qubit readout. This large difference is additional proof of the effectiveness of the quantum filter.}
\end{figure}

\subsection{Statistics of photon jumps}

In order to test how faithfully our repeated parity measurement can track photon losses, we simply count the number of jumps extracted from the parity estimator during 500~$\mu$s repeated parity measurements. We have applied a Schmitt trigger to digitize the parity estimator to reject the unavoidable noise (spikes in the estimator) coming from qubit decoherence and erred parity readout. The two thresholds for the Schmitt trigger are chosen to be $\pm0.9$ for a large discrimination. Then the number of parity jumps is inferred from the number of transitions in the digital data after the Schmitt trigger. 

\begin{figure}
\includegraphics{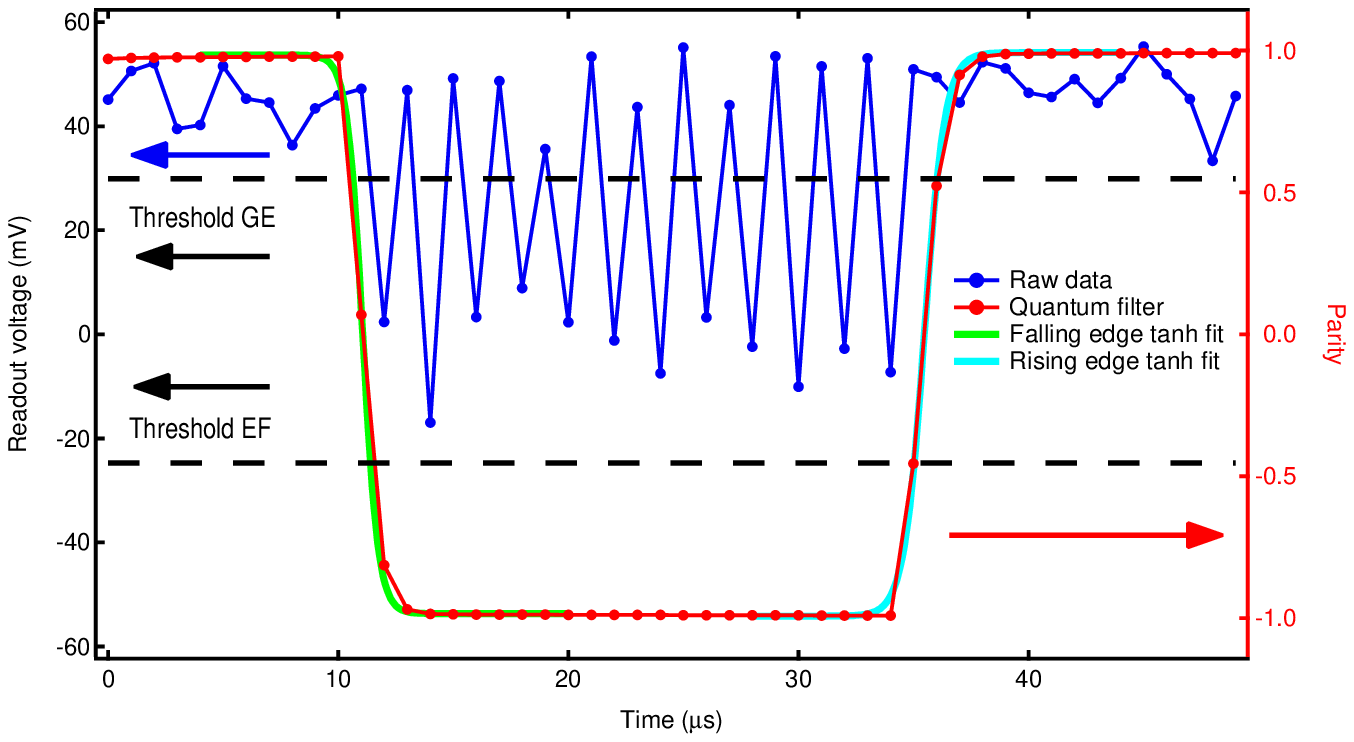}
\caption{Response time of the quantum filter applied to typical photon jump events. Blue curve is the raw data from a repeated parity measurement. Red curve is the corresponding parity estimator based on the quantum filter. Green (cyan) curves are fits to tanh functions of the parity estimator at the transition down (up), giving a transition time constant of less than 1~$\mu$s. However, the response time of the filter to make a transition between $\pm0.9$ is $\tau_f\sim2~\mu$s.}
\end{figure}

Although our averaged single parity readout fidelity is 80\% (90\% to be correct and 10\% to be wrong), due to the averaging effect of the quantum filter, we actually can achieve nearly unity detection sensitivity of single photon jump events. However, because of the finite bandwidth of the filter, if two photon jumps occur within the response time of the filter $\tau_f$ (defined as the time to make a transition between the two thresholds for the Schmitt trigger), our Schmitt trigger will not catch both jumps. Figure~S8 shows the time response of the quantum filter applied to typical photon jump events. Green and cyan curves are fits of the parity estimator at the transition based on a tanh function, giving a transition time constant less than 1~$\mu$s. We also find the response time of the filter to make a transition between $\pm0.9$ is $\tau_f\sim2~\mu$s. The probability of having a second photon jump within $\tau_f$ after the first jump is simply $\mathbb{P}_{jump}=\frac{\bar{n}}{\tau_{\rm tot}}\int_0^{\tau_f}\mathrm{e}^{-t\bar{n}/\tau_{\rm tot}}\,\mathrm{d}t=1-e^{-\tau_f\bar{n}/\tau_{\rm tot}}$. For $\bar{n}=1$ and $\tau_{\rm tot}=49~\mu$s, the above probability is $\mathbb{P}_{jump}=4\%$, while $\mathbb{P}_{jump}=15\%$ for $\bar{n}=4$, which is the probability of missing both jumps. 

Figure~S9 shows the histograms of the extracted number of jumps for an initial even or odd cat state by post selections. We note that the almost non-mixing distribution of even and odd numbers is trivial due to the following reason. At the end of 500~$\mu$s repeated parity measurements, the cavity is already in a steady state with $n_{th}=0.02$ photons, that is 98\% probability at vacuum (even parity) and 2\% probability with one photon (odd parity). When the initial parity of the cat state, for example an even parity, is determined by post selections, the number of jumps should have 98\% probability of being even and only 2\% probability of being odd closely tied with the distribution of the final parity at $t=500~\mu$s. Similar argument applies to an initially odd parity cat. The even/odd distributions in Fig.~S9 indeed show a 98\%-2\% mixing, providing another way of determining $n_{th}$. 

In reality, we have no way of knowing the true number of photon jumps for each parity measurement trajectory. The only way to test how faithfully our repeated parity measurement can track photon jumps is to see whether the distribution of jumps agrees with what we expect. Due to the complication of background thermal excitation and finite response time of the filter, to get an analytic solution is difficult. Instead, we perform a numerical Monte Carlo simulation to compare with the experiment. In the simulation, we use a coherent state as the initial state without distinguishing the parity.  Each simulation trajectory is 500~$\mu$s long including a transition probability $n \rightarrow n+1$ from the background thermal excitation. In the simulation, we also neglect those who have neighboring jumps within the response time $\tau_f$ of the quantum filter. Then for each trajectory we count the number of jumps and finally we perform a histogram (black solid lines in Fig.~S9) of those numbers based on 100,000 trajectories. The good agreement between simulation and data demonstrates that the repeated parity measurement can track the error syndromes faithfully.

\begin{figure}
\includegraphics{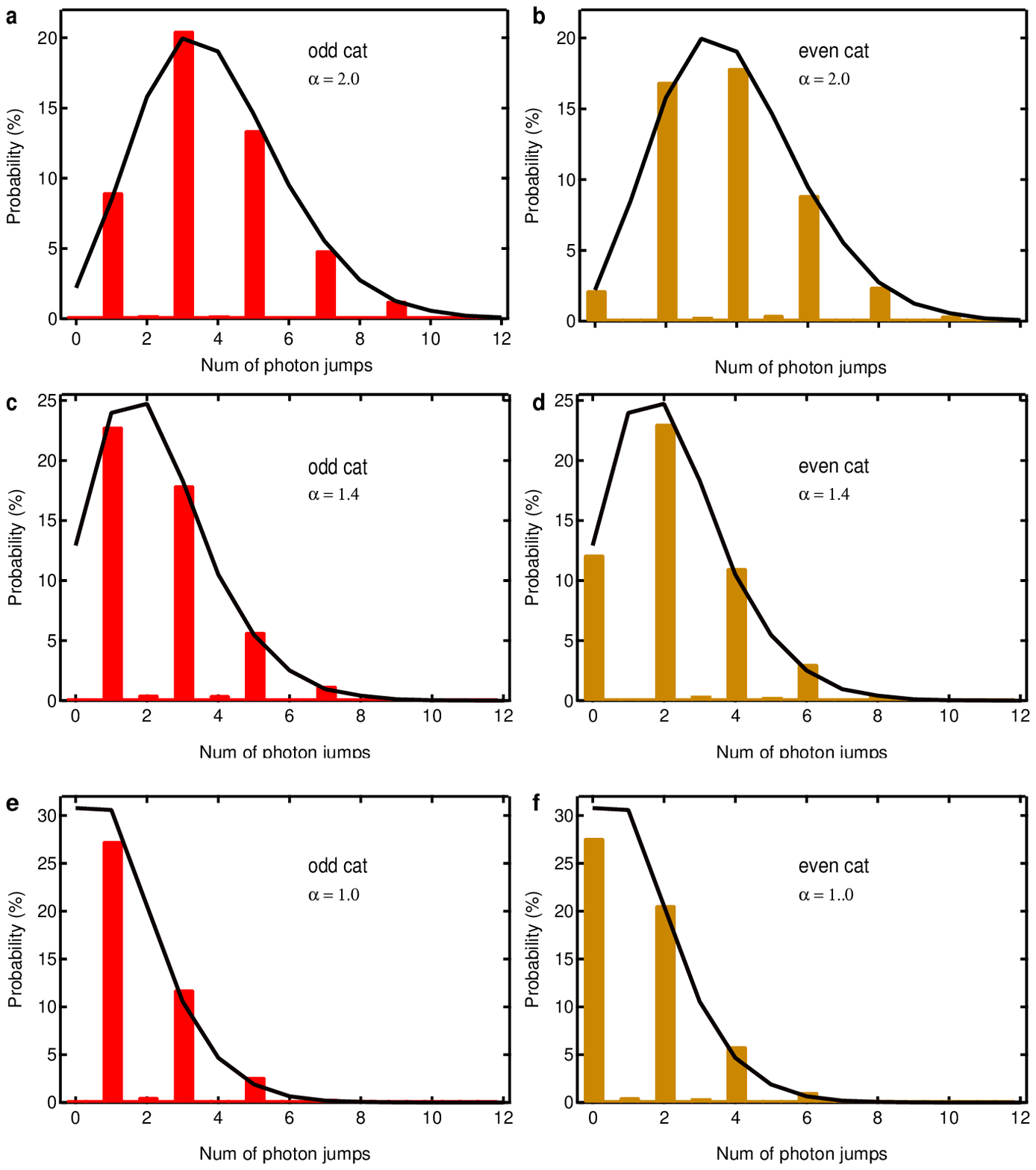}
\caption{Histograms of the number of jumps extracted from the parity estimator during 500~$\mu$s repeated parity measurements for an initial even or odd cat state by post selections. (a) and (b) are for $|\alpha|=2.0$, (c) and (d) are for $|\alpha|=1.4$, and (e) and (f) are for $|\alpha|=1.0$. Solid lines are numerical simulations including the background thermal excitation and finite response time of the quantum filter. In the simulation, we use a coherent state as the initial state without distinguishing the parity. The good agreement between data and simulation demonstrates that the repeated parity measurement can track the error syndromes faithfully.}
\end{figure}

\subsection{Quantifying parity tracking performance}

Our demonstrated parity tracking protocol has two major sources of infidelity that lead to the decay of our cat states, ultimately putting a bound on the improvement we would be able to achieve in an actual QEC protocol. These are: missing photon jumps and qubit $T_1$ decay. Missing a jump would result in an errant interpretation of the cavity state we are decoding. Qubit $T_1$ decay would instead impart an arbitrary phase on the cat states that without some auxiliary correction protocol would be impossible to recover from.

Given our system's parameters, we can quantify what level of improvement we can achieve with the demonstrated parity tracking protocol over a photon jump rate $\bar{n}\kappa$. An optimal balance must be struck between the infidelity induced by each of the two mechanisms. If one measures too frequently, qubit $T_1$ decay will dominate the decay due to missing jumps. Conversely, not measuring often enough, although mitigating the effects of qubit errors, risks missing a photon jump. In particular, one can write down an effective decay rate $\kappa_{eff}$ as:
\begin{equation}
\kappa_{eff}=[\frac{(\bar{n}\kappa)^2(\tau_M+\tau_W)^2}{2}+P_C(T_1)]\frac{1}{\tau_M+\tau_W},
\end{equation}
where $\tau_M$ is the parity measurement time, $\tau_W$ is a waiting time between two consecutive parity measurements that can take on any value $\geq0~\mu$s, and $P_C(T_1)$ is a constant    probability of dephasing due to qubit decay. The minimum $\kappa_{eff}$ is achieved when the decay rates are equal:
\begin{equation*}
(\tau_M+\tau_W)^2=\frac{2P_C(T_1)}{(\bar{n}\kappa)^2}\Rightarrow\kappa_{eff}=\bar{n}\kappa\sqrt{2P_C(T_1)}
\end{equation*}
The improvement over $\bar{n}\kappa$ is thus on the order of $\sqrt{2P_C(T_1)}$.  The probability $P_C(T_1)$ can simply be taken as $\tau_M/T_1$, the relevant figure of merit that quantifies a worst case scenario, namely that the qubit remains in $\ket{e}$ during and after each measurement.  However, with current real-time feedback technologies rapidly advancing, it should in principle be possible to keep the qubit in $\ket{g}$ after each measurement. Given each full parity measurement takes $1~\mu$s in our system, $\tau_M$ could in principle be cut down to $\sim400$~ns, with $\pi/\chi_{qs}\sim275$~ns and a projective measurement lasting just over $\sim100$~ns.  Putting the numbers together, one could in principle enhance the lifetime of a quantum bit encoded in the resonator by a factor of 3, from $1/\bar{n}\kappa=12~\mu$s to $\sim36~\mu$s.  In addition, the optimal waiting time between measurements $\tau_W$ would be $\sim4~\mu$s.  Given that $\tau_M$ is dominated in large part by the parity protocol waiting time $t=\pi/\chi_{qs}$, a relevant benchmark for the overall performance becomes the product $\chi_{qs}T_1$.  We emphasize that even for this system's modest coherence properties, a factor of 3 improvement would be significant.  Indeed, just doubling $T_1$ would already take the lifetime of the information to $52~\mu$s, the lifetime of a single photon Fock state in the storage cavity.